\documentclass[11pt,a4paper]{article}
\pdfoutput=1
\usepackage{jheppub}

\begin{document}

\noindent
\title{Closing in on a perturbative fourth generation}
\author[a]{Mathieu Buchkremer, Jean-Marc G\'{e}rard and Fabio Maltoni}
\affiliation[a]{Centre for Cosmology, Particle Physics and Phenomenology (CP3),\\
Universit\'{e} catholique de Louvain,\\
Chemin du Cyclotron, 2, B-1348, Louvain-la-Neuve, Belgium.}
\emailAdd{mathieu.buchkremer@uclouvain.be}
\emailAdd{jean-marc.gerard@uclouvain.be}
\emailAdd{fabio.maltoni@uclouvain.be}
\abstract{A perturbative new family of fermions is now severely constrained, though not excluded yet. We reconsider the current bounds (i.e., direct and from Higgs searches, $R_b$, oblique parameters) on the fourth generation parameter space assuming the case of a small CKM mixing with the third generation. We identify viable scenarios featuring either a light or a heavy Higgs boson. A set of representative benchmark points targeted for LHC searches is proposed with a normal (inverted) quark mass hierarchy where $t^{\prime }\rightarrow b^{\prime }W$ ($b^{\prime }\rightarrow t^{\prime }W$%
) are sizable. In the case where the fourth generation couplings to the lighter quark families are small, we suggest that search strategies at the LHC should include both pair (strong) and single (weak) production with $bb+nW$ ($n=2,...,6$) final state signatures.}

\arxivnumber{1204.5403}
\maketitle

\section{Introduction}

\hspace{0.75cm} It is generally claimed that the Standard Model (SM) comprises three generations of fermions.
However, many fundamental problems do not find an answer into this framework, and the possibility of
additional massive fermions, such as a new sequential family of quarks, is currently among the models in the
spotlight of experimental searches at the LHC (for an overview, see, e.g., \cite{1}). Besides its phenomenological
relevance, the Standard Model with a fourth generation (hereafter SM4) can also serve as a template for
new physics models for which the unitarity of the $3\times 3$ CKM matrix might be violated \cite{2}.
Despite the absence of hints in experimental searches, there is currently a renewed interest in the fact that new fermion families
are still allowed by the electroweak precision constraints if additional parameters are considered.\ For instance,
considering non-zero mixings between a new heavy doublet and the lighter
quark generations has recently been shown to widen the allowed parameter space \cite{3}. Furthermore, the mass difference between the new heavy quarks also requires specific attention. In their seminal paper \cite{4}, Kribs \emph{et al.}\ proposed the correlation
\begin{equation}
\Delta m_{q^{\prime }}\simeq \Big(1+\frac{1}{5}\ln \frac{m_{H}}{115\text{ GeV}}%
\Big)\times 50\text{ GeV} \label{1}
\end{equation}%
between the SM4\ Higgs boson mass and the quark mass difference, $\Delta m_{q^{\prime }}=m_{t^{\prime }}-m_{b^{\prime }}$. While simple, this approximation
is known to overconstrain the fourth generation parameter space \cite{5}%
, and the experimental searches now require more refined predictions. 

\newpage

\begin{equation*}
\begin{tabular}{|c||c|c|c|c|c|c|}
\hline
& $m_{q^{\prime }}$ & \multicolumn{1}{|c|}{$\Delta m_{q^{\prime }}$} & $%
s_{34}$ & $m_{l^{\prime }}$ & $\Delta m_{l^{\prime }}$ & $m_{H}$ \\ 
\hline\hline
Direct searches & $\checkmark $ &  &  & $\checkmark $ &  &  \\ \hline\hline
Higgs searches &  & \multicolumn{1}{|c|}{} &  &  &  & $\checkmark $ \\ \hline
$Z\rightarrow b\bar{b}$ & $\checkmark $ & \multicolumn{1}{|c|}{$\checkmark $}
& $\checkmark $ &  &  &  \\ \hline
$S,T$ & $\checkmark $ & \multicolumn{1}{|c|}{$\checkmark $} & $\checkmark $
& $\checkmark $ & $\checkmark $ & $\checkmark $ \\ \hline\hline
Strong production & $\checkmark $ &  &  &  &  &  \\ \hline
Electroweak production & $\checkmark $ & $\checkmark $ & $\checkmark $ &  & 
&  \\ \hline
Branching Ratio &  & $\checkmark $ & $\checkmark $ &  &  &  \\ \hline
\end{tabular}%
\end{equation*}%
Table 1: Summary of the relevant fourth generation parameters for the LHC searches,
with the fourth generation fermions $q^{\prime }=t^{\prime },b^{\prime }$
and $l^{\prime }=\tau ^{\prime },\nu ^{\prime }$ ($\Delta m_{l^{\prime
}}=m_{\nu ^{\prime }}-m_{\tau ^{\prime }}$).\ The marks denote the
dependencies on the constraints considered in our analysis. 

\bigskip

In this work we argue that a better description of the fourth generation parameter
space is provided by
\begin{equation}
\Delta m_{q^{\prime }}\lesssim \Big(1+\frac{1}{5}\ln \frac{m_{H}}{125\text{ GeV}}%
-15\text{ }s_{34}^{2}\Big)\times m_{W}, \label{2}
\end{equation}%
where $s_{34}\equiv \sin \theta _{34}=|V_{t^{\prime }b}|$ is the CKM mixing element between the two heaviest quark families.

\smallskip

In the following, we scrutinise the various SM4 scenarios
uncovered by the current approximations, consistently with the upper limit (\ref{2}). In Section 2, we
review the current bounds on the relevant fourth generation parameters in collider searches, taking into account the possibility for small but non-zero CKM mixings with the third family. Section 3\ describes our numerical analysis of the fourth generation parameter space, allowed by the electroweak precision observables.
Considering the relative importance of the quark mass splittings, we present a selection of representative benchmark points for the
normal ($m_{t^{\prime }}>m_{b^{\prime }}$) and inverted ($m_{t^{\prime
}}<m_{b^{\prime }}$) quark mass hierarchies with a particular focus on the
light Higgs scenario. In Section 4, we detail the production and
experimental signatures corresponding to these scenarios, and work out the
associated cross-sections and decay rates. Our conclusions are presented in
Section 5.

\section{Current bound dependencies on the fourth generation parameters}

While strong pair production at the LHC only depends on the new heavy quark masses, the electroweak production and decay modes are mainly sensitive to their mass difference and CKM mixings. As our purpose is to provide representative benchmark scenarios, the relations between the various SM4 parameters must be investigated carefully. Following Table 1, we first analyse the direct limits on the fourth generation fermion masses.\ The Higgs direct hints provide an
independent constraint on the allowed parameter space due to the decoupling
of the new heavy fermion masses.\ The electroweak precision parameters
are then considered, with the non-oblique correction to the $Z\rightarrow b%
\bar{b}$ vertex setting a stringent upper bound on $s_{34}$ as a function of
the quark masses. Finally, the oblique parameters $S$ and $T$ are used to
derive (\ref{2}) as a function of all previous constraints.\ 

\newpage

\subsection{Direct limits}

CDF\ and D0 experiments previously set the strong limits $m_{t^{\prime }}>335$
GeV \cite{6} and $m_{b^{\prime }}>385$ GeV \cite{7} at the 95\% confidence level (CL) for the
masses of new up- and down-type heavy quarks. Yet, direct searches at
the LHC now hint at even more stringent constraints. Searching for short-lived $%
b^{\prime }$ quarks in the signature of trileptons and same-sign dileptons,
CMS\ ruled out $m_{b^{\prime }}<611$\ GeV\ at 95\% CL assuming BR$(b^{\prime
}\rightarrow tW)\simeq 100$ $\%$ \cite{8}.\ A search for heavy pair-produced
top-like quarks has also been conducted considering $%
t^{\prime }\rightarrow bW$ as a prompt exclusive decay.\ No excess over the
SM expectations has been observed, excluding a $t^{\prime
}$ quark with a mass below $557$ GeV \cite{9}. The inclusive search \cite{10}
also set a strong limit on a degenerate fourth generation, $%
m_{t^{\prime }}>495$ GeV at 95\%\ CL, assuming a minimal off-diagonal mixing
between the third and the fourth generation.\ Recently, the ATLAS\
collaboration ruled out $m_{Q}<350$ GeV at 95\%\ CL by searching for
pair-produced heavy quarks $Q\bar{Q}$ in the two-lepton channel \cite{11},
with BR$(Q\rightarrow qW)\simeq 100$ $\%$, where $q=u,d,c,s,b$. This last
analysis does not rely on b-tagging and provides a more conservative bound, which we will use in the following.

\bigskip

On the other hand, we observe that most of the aforementioned limits are model-dependent, so that some regions of the parameter space are still allowed and the associated (peculiar) phenomenology unexplored. 
Firstly, direct searches rely on the hypothesis
that new heavy quarks are close in mass and mix predominantly with the third generation, which implies short lifetimes and prompt decays in the beampipe. While the former condition is well motivated by the precision electroweak measurements, a small yet non-zero 3-4 family CKM mixing element must be considered very carefully. Secondly, the current limits now reach the bound 
$m_{t^{\prime }}\lesssim 550$ GeV required from the unitarity for the partial
S-wave amplitude in $t^{\prime }\bar{t}^{\prime }$ scattering at tree-level 
\cite{12}. Additionally, perturbative methods could become inadequate. As shown in \cite{13} from the 2-loop renormalisation group equations, the fourth generation Yukawa couplings lead to the breakdown of perturbativity below the TeV scale if $m_{t^{\prime },b^{\prime }}>375$ GeV. In summary, with the forthcoming searches pointing at even larger mass scales, the possibility for long-lived heavy fermions should be considered, as well as a strongly coupled fourth generation or Yukawa bound states \cite{14,15,16,17}. 

\bigskip

Regarding the fourth generation lepton sector, the LEP obtained the lower bound $m_{\tau ^{\prime }}>100$ GeV for new heavy charged leptons \cite{18}.\ Assuming Dirac masses, the $Z$ invisible width set $m_{\nu ^{\prime }}>45$ GeV for new heavy stable neutrinos \cite{19}.\ Although the off-diagonal lepton mixings  are required to be smaller than 0.115 to be consistent with the recent global fits  \cite{20}, a trivial $4\times 4$ PMNS\ unitary matrix is considered throughout this analysis, namely a long-lived Dirac neutrino with $|U_{\nu ^{\prime }\tau ^{\prime
}}|=1$. We mention, however, the interest of the Majorana case, as the exhaustive study \cite%
{21} has shown.

\newpage

\subsection{Higgs search constraints}

The recent results from the CMS\ experiment (with integrated luminosity of $4.8$ fb$^{-1}$) now exclude a SM Higgs boson mass between $127.5$ GeV and $600$ GeV  at 95\%\ CL with all channels combined \cite{22}. A local significance of 3.1$\sigma$ is obtained for a $124$ GeV SM Higgs
boson decaying to two photons \cite{23}. ATLAS observed an excess of events around $m_{H}=125$ GeV ($126.5$ GeV) with a local significance of 2.5$\sigma$ (2.8$\sigma $) in $H\rightarrow ZZ^{\ast }\rightarrow 4l$ ($H\rightarrow \gamma \gamma $), constraining $m_{H}$ in the ranges $117.5-118.5$ GeV and $122.5-129$ GeV with an integrated luminosity of $4.9$ fb$^{-1}$ \cite{24,25,26}. As far as a fourth family of fermions is considered, CMS recently
excluded a SM4 Higgs boson in the mass window 120-600 GeV at 95\% CL \cite{27},
accounting for $\sim 600$ GeV heavy fermion masses and in agreement with the spectrum (\ref{1}). Under similar assumptions,
ATLAS\ rules out a fourth generation SM Higgs boson with mass
between $119$ and $593$ GeV at the two sigma level \cite{28}. 

While very little room seems to be left for a light
SM4\ Higgs boson, recent studies emphasised that the above exclusion ranges
did not take into account the possibility for a stable fourth generation neutrino \cite{29,30}.\ If such new neutral leptons are lighter than half the Higgs boson mass, the opening of the new invisible mode $%
H\rightarrow \nu ^{\prime }\bar{\nu}^{\prime }$ increases the Higgs
total width, and overtakes the $H\rightarrow WW^{(\ast )},ZZ^{(\ast )}$ and $f%
\bar{f}$ rates with a substantial branching ratio in the low mass region. However, it is straightforward to check that the recent hints of a $125$ GeV scalar boson fine tune $m_{\nu ^{\prime }}$ to a value very close to half the Higgs boson mass. Given that $H\rightarrow b\bar{b}$ is the dominant visible signal for a light SM-like Higgs boson with respect to $H\rightarrow \nu
^{\prime }\bar{\nu}^{\prime }$ in SM4, the ratio 
\begin{equation}
\frac{\Gamma _{H}^{SM4}}{\Gamma _{H}^{SM}}\simeq \frac{\Gamma (H\rightarrow
\nu ^{\prime }\bar{\nu}^{\prime })_{SM4}}{\Gamma (H\rightarrow b\bar{b})_{SM}%
}\simeq \frac{m_{\nu ^{\prime }}^{2}}{3m_{b}^{2}}\Big(1-4\frac{m_{\nu ^{\prime
}}^{2}}{m_{H}^{2}}\Big)^{3/2} \label{12}
\end{equation}%
must be of order unity. Combining (\ref{12}) with the assumption that the Higgs invisible branching ratio does not excess 50\%, the new neutral lepton mass must lie between 61\ and 62.5\ GeV. Were direct searches to exclude $m_{\nu ^{\prime }}\lesssim m_{H}/2$,
the Higgs decay patterns would not be affected by the neutrino mass, and the
aforementioned exclusion ranges apply. If $m_{\nu ^{\prime }}>m_{H}/2$, a light SM4\ Higgs boson is expected to lead to significantly more events. In particular, the $H\rightarrow f\bar{f}$ and $H\rightarrow ZZ^{(\ast )}$ signal strengths are enhanced by a factor 5 to 6 for $m_{H}\simeq 125$ GeV, given the gluon fusion enhancement \cite{31}.\ This remains consistent with
the recent LHC\ results, except in the channel $H\rightarrow \tau \tau $, for which the excess observed by CMS\ is currently inconsistent with the SM4 predictions \cite{27}, but is still to be confirmed by the
ATLAS\ results.

A new family of fermions implies important consequences for
collider phenomenology, as it significantly modifies the Higgs production and decay rates with respect to SM. At the leading order, the Higgs production cross-section via gluon-gluon fusion increases by a factor of about 9 due to the additional $t^{\prime }$ and $b^{\prime }$ fermion loops arising from the decoupling of the fourth generation quark masses \cite{4}. The same enhancement in $H\rightarrow gg$ increases the Higgs total width by a factor of about 1.6. On the other hand, the $H\rightarrow \gamma \gamma $ width suffers a factor of about 5 suppression due to the new heavy fermion loops involving $t^{\prime },b^{\prime }$ and $\tau ^{\prime }$. This accidental cancellation against the $W$ contribution in the $H\rightarrow \gamma \gamma $ amplitude results in a comparable number of diphoton events.

\newpage
Recently, the calculation of the NLO electroweak corrections hinted at very large radiative corrections to the Higgs decay amplitudes in SM4. Considering fourth generation fermion masses larger than 450 GeV (600 GeV), the Higgs
branching ratios into $WW$ and $ZZ$ pairs are suppressed by corrections of
the order of -40\% (-60\%), whereas BR$(H\rightarrow f\bar{f})$ is
enhanced by 30\% (60\%) in SM4 \cite{32}. Additionally, the rate $%
\sigma (gg\rightarrow H)\times $ BR$(H\rightarrow \gamma \gamma )_{|SM4}$ is
further suppressed by one order of magnitude with respect to the LO result, which hardly accomodates the fourth generation fermion scenario \cite{33,34,35}. 
However, such an accidental difference between the LO and NLO corrections raises legitimate
concerns about the stability of the above predictions. Therefore, this, together with the fact that the $\gamma \gamma $ signal might still be a statistical fluctuation, suggest us to leave open the light Higgs window between $117$ GeV and $135$ GeV, i.e., up to where the decay mode into a $b\bar{b}$ pair is dominant.

\bigskip

Finally, we emphasise that if the hints for a SM-like light scalar signal would be due to statistical fluctuations and disappear with more integrated luminosity at the LHC, the possibility for a heavy Higgs would need to be reconsidered. Although the CMS and ATLAS experiments rule out $m_{H}<600$ GeV in the context of SM4, it has been noted in \cite{36} that the use of the narrow width approximation
to set exclusion limits on a SM Higgs boson should be avoided for larger mass values. In particular, the effects due to off-shell production and decay can lead to large deviations with respect to the zero-width approximation for Higgs boson masses larger than $300$ GeV. Given that a proper description of the Higgs line shape is required to set more accurate exclusion limits at such scales, the conservative lower bound $m_{H}>450$ GeV will be considered in the following for the high mass region.

\subsection{Electroweak precision observables}

The electroweak precision observables are well known to yield stringent limits on the possible extensions of the SM, and set up compelling bounds on the allowed mixings between a sequential fourth generation and the lighter fermion families. Considering a generic flavour structure, the addition of a new quark doublet requires to extend the CKM\ mixing sector to a $4\times 4$ unitary matrix, whose entries depend upon three additional mixing angles and on two new
CP-violating phases. Assuming a non-zero 3-4 family CKM mixing, the $R_{b}$ ratio $\Gamma (Z\rightarrow b\bar{b})/\Gamma (Z\rightarrow $
hadrons$)$ can be shown to provide a solid constraint on a non-degenerate fourth generation, independently of the Higgs and leptonic sectors. On the one hand, the non-oblique corrections to $R_{b}$ are known to remain below several permille for $m_{\nu ^{\prime }}<100$ GeV \cite{37}. On the other hand, the nondecoupling effects present in the $Zb\bar{b}$ vertex turn out to be slightly sensitive to non-zero $m_{t^{\prime }}-m_{b^{\prime }}$ quark mass splittings due to
non-negligible $\mathit{O}(\alpha _{s}^{2})$ QCD\ corrections. Following the derivation given in \cite{38} and \cite{39}, their effect on the $Z$ decay rate depends on the two quantities $\delta _{b}$ and $\delta _{tQCD}^{q}$. While vanishing for light quarks, $\delta _{b}$ is sensitive to $t$ and $%
t^{\prime }$ loop corrections in $Z\rightarrow b\bar{b}$, and reads
\begin{equation}
\delta _{b}\simeq \Big[ \Big(0.2-\frac{m_{t}^{2}}{2m_{Z}^{2}}\Big)\text{ }%
|V_{tb}|^{2}+\Big(0.2-\frac{m_{t^{\prime }}^{2}}{2m_{Z}^{2}}\Big)\text{ }%
|V_{t^{\prime }b}|^{2}\text{ }\Big]\times 10^{-2}.
\end{equation}

\newpage

Conversely, $\delta _{tQCD}^{q}$ includes the QCD\
contributions to the axial part of the decay. As computed in \cite{40}, the
induced axial and vector currents corrections to the Z\
boson decay rate receive significant $\mathit{O}(\alpha _{s}^{2})$ corrections due to the
possible large splitting of heavy quark doublets.\ They stem from the
absorptive part of the diagram obtained by squaring the tree-level decay $%
Z\rightarrow q\bar{q}$ with the real emission of 2\ gluons $Z\rightarrow q%
\bar{q}\rightarrow gg$ via a triangle loop. In the context of four quark generations, the incomplete
cancellation between the various heavy doublets gives rise to%
\begin{eqnarray}
\delta _{tQCD}^{f} &=&-\frac{\hat{a}_{f}}{\hat{v}_{f}^{2}+\hat{a}_{f}^{2}}\Big(\frac{\alpha _{s}}{\pi }\Big)^{2}[\hat{a}_{t}\text{ }f(\mu _{t})+\hat{a}_{t^{\prime }}\text{ }f(\mu
_{t^{\prime }})-\hat{a}_{b^{\prime }}\text{ }f(\mu _{b^{\prime }})]
\end{eqnarray}%
where $\alpha _{s}$ is the QCD coupling constant evaluated at the Z pole (i.e. $\alpha _{s}(M_{Z}^{2})=0.119$) and $a_{q}=2I_{3}^{q}$, $%
v_{q}=2I_{3}^{q}-4Q_{q}\hat{s}_{W}^{2}$ are the vector and axial vector
couplings of the $q$ quark, respectively. The quark masses $\bar{m}%
_{i}$ are defined in the $\overline{MS}$ scheme, with $\mu _{i}^{2}=4\bar{m}_{i}^{2}/M_{Z}^{2}$.

\bigskip

The function $f$ reads \cite{39}%
\begin{equation}
f(\mu _{i})=\text{ln }\Big(\frac{4}{\mu _{i}^{2}}\Big)-3.083+\frac{0.346}{\mu
_{i}^{2}}+\frac{0.211}{\mu _{i}^{4}}.\text{ }
\end{equation} 
The $\delta _{tQCD}^{f}$ correction depends explicitely on \emph{both} fourth generation quark masses, providing an additional constraint on the 3-4 family mixing as a function of the mass splitting. Using the exact expressions given in \cite{38}, we obtain
\begin{equation}
R_{b}=\frac{1.594(1+\delta _{b})}{5.722+1.594(1+\delta _{b})}\label{20}
\end{equation}
for $m_{t^{\prime }}=500$ GeV, which we can compare to the experimental value $R_{b}^{\exp }=0.21629\pm 0.00066$ \cite{19}. We estimate that $\delta _{tQCD}^{f}$ yields a $-5\%$ ($+4\%$%
) correction for $m_{t^{\prime }}=500$ GeV if $\Delta m_{q^{\prime }}=80$ GeV ($%
-80$ GeV), which effect becomes weaker for larger $m_{t^{\prime }}$ masses. While this implies a
measurable effect in $R_{b}$, it depends sensitively on our choice of the $\overline{MS}$ top quark mass, here taken as $\bar{m}_{t}=163.5$ GeV. Combining (\ref{20}) with the condition $R_{b}>R_{b}^{\exp }$
at the $2\sigma $ level, we obtain the following upper bound on the 3-4 CKM mixing (for $\Delta m_{q^{\prime }}=0$ GeV) :%
\begin{equation}
(s_{34})_{R_{b}}\leq \left\{ 
\begin{array}{c}
0.17\text{ if }m_{t^{\prime }}=400\text{ GeV} \\ 
0.13\text{ if }m_{t^{\prime }}=500\text{ GeV} \\ 
0.11\text{ if }m_{t^{\prime }}=600\text{ GeV}%
\end{array}%
\right.
\end{equation}%
which can be approximated by
\begin{equation}
\text{ }(s_{34})_{R_{b}}\lesssim 0.8\text{ }\frac{m_{W}}{m_{t^{\prime }}}. \label{10}
\end{equation}%
Additionally, such bounds occur to be sensitive to the quark mass
splitting when accounting for the $\mathit{O}(\alpha _{s}^{2})$ QCD corrections in the $%
Z\rightarrow b\bar{b}$ decay rate. Most important, larger quark masses favour smaller $t-b^{\prime }$ and $t^{\prime }-b$ mixings, independently of the Higgs boson and heavy neutrino masses.

\newpage

Interestingly, the oblique parameters originally introduced by Peskin and Takeuchi \cite{41} lead to similar, albeit less model independent conclusions. Including all
one-loop self-energy contributions of the electroweak vector bosons $W, Z$ and $\gamma $ arising from new heavy chiral fermions, the $S$, $T$ and $U$
have been evaluated in the presence of a perturbative fourth family of fermions in \cite%
{42}. Assuming that the corresponding new physics
scale is much larger than the electroweak scale and that additional
couplings to the SM fermions are suppressed, the contributions from a fourth fermion family to the oblique parameters read
\begin{eqnarray}
S_{4} &=&\frac{1}{3\pi }\Big(2+\ln \frac{m_{b^{\prime }}\text{ }m_{\nu ^{\prime
}}}{m_{t^{\prime }}\text{ }m_{\tau ^{\prime }}}\Big),  \label{3} \\
T_{4} &=&\frac{3\text{ }[m_{t^{\prime }}^{2}+m_{b^{\prime
}}^{2}+s_{34}^{2}(F_{t^{\prime }b^{\prime }}-F_{tb^{\prime }})-F_{t^{\prime
}b^{\prime }}]+[m_{\tau ^{\prime }}^{2}+m_{\nu ^{\prime }}^{2}-F_{\tau
^{\prime }\nu ^{\prime }}]}{16\pi \text{ }s_{W}^{2}\text{ }c_{W}^{2}\text{ 
}M_{Z}^{2}},
\end{eqnarray}%
where $s_{W}$\ and $c_{W}$\ define the sine and cosine of the weak mixing angle respectively, and

\begin{equation}
F_{ij}=\frac{2m_{i}^{2}m_{j}^{2}}{m_{i}^{2}-m_{j}^{2}}\ln\frac{%
m_{i}^{2}}{m_{j}^{2}}.
\end{equation}

We observe that large quark mass differences increase the $T_{4}$
contribution, while small $s_{34}$ values balance this effect. Assuming $|m_{\tau ^{\prime }}-m_{\nu ^{\prime }}|<90$ GeV and $m_{t^{\prime }}\simeq 500$ GeV, these observations
can be summarised by the approximate upper bound (\ref{2}) which the $T$ parameter constraint translates to if $T\lesssim 0.3$ is fulfilled. As depicted in Figure 1, it extends the
approximation (\ref{1}) and includes the mixing angle dependence
between the two heavier generations, setting an upper limit
on the allowed quark mass splitting from the $T$ parameter. 

\bigskip

\begin{figure}[htbp]
\centering\includegraphics[scale=0.68]{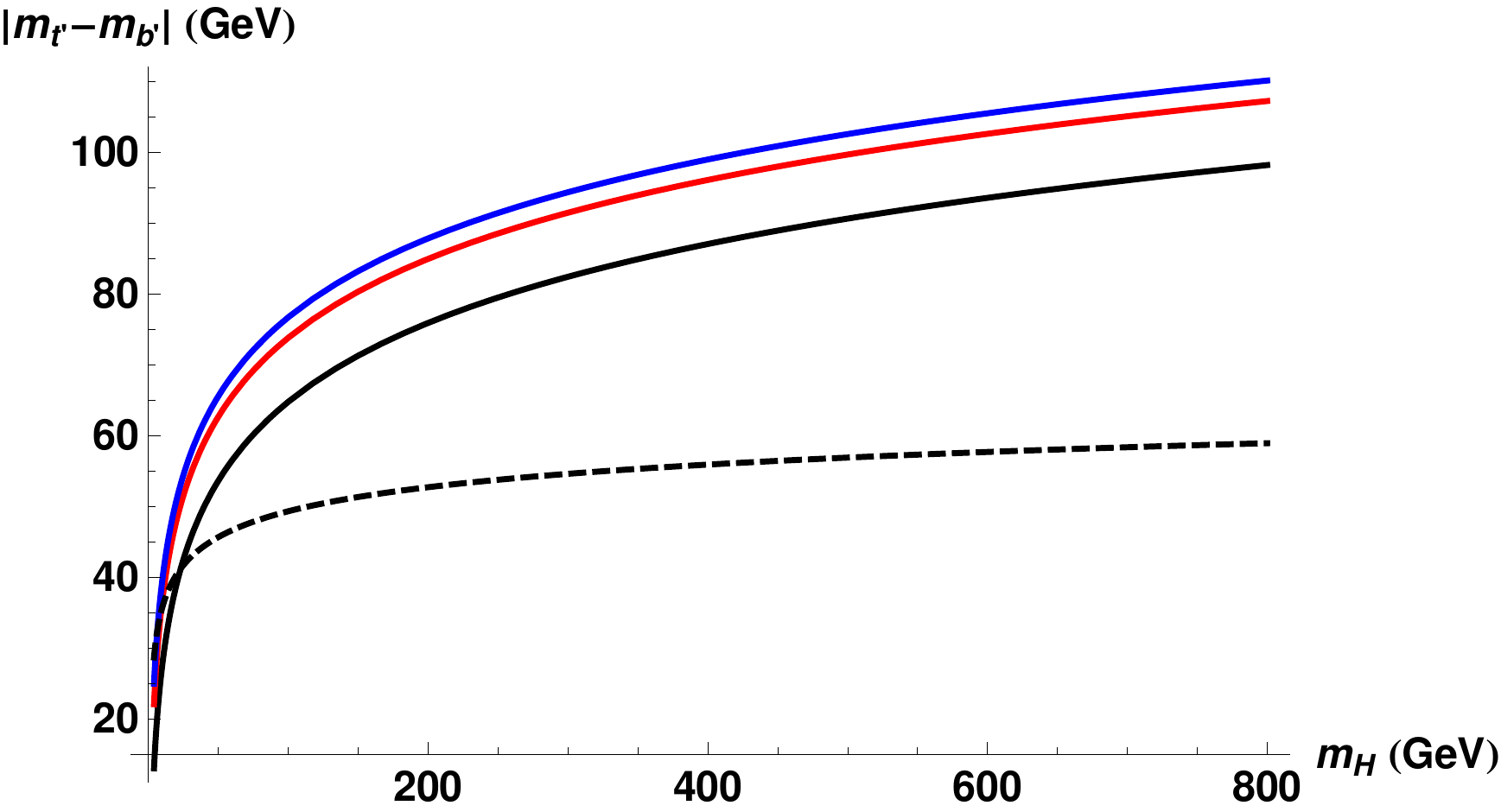}
\caption{95\%\ CL\ upper bound on $|\Delta m_{q^{\prime }}|$ versus $m_{H}$
for $m_{t^{\prime }}=500$ GeV as given in (\protect\ref{2}) from the $T$ parameter constraint, for the mixing values $s_{34}=0.1$ (black), $0.05$ (red) and $0.01$ (blue). The dashed line
depicts the value of the quark mass splitting as defined by the relation (%
\protect\ref{1}). No significant change is observed below $s_{34}<10^{-2}$.}
\label{}
\end{figure}

\newpage

The corrections to the $U_{4}$ parameter are known to be negligible if $%
T_{4}<0.4 $ \cite{42}, so that we restrict our analysis to the two
dimensional $S-T$\ parameter space in the following.\ 
The $S_{4}$ parameter is sensitive to the mass
scale of the new chiral fermions, whereas $T_{4}$ constrains the mass
splitting between all isospin partners. Accordingly $T_{4}$
increases with $s_{34}$, while $S_{4}$ is CKM-independent.\ Besides the fermion contributions, a heavy Higgs boson also
affects the oblique corrections, which are known to depend logarithmically on the Higgs mass. In our analysis, the complete corrections computed at the $Z$ pole given in \cite{43} have been used. If larger than the SM\ reference, the Higgs boson mass $m_{H}$
simultaneously lowers down the $T$ parameter and increases $S$. On the other hand, a smaller 3-4\ mixing reduces the $T$
parameter for a heavy Higgs mass without prejudicing the $S$ parameter.

\bigskip

While small values are favored in the light
Higgs case, fermion mass splittings larger than the $W$ mass remain allowed in significant regions of the SM4 parameter space if $m_{H}\geq 450$ GeV.\ In the presence of a finite 3-4 CKM mixing element, the nondecoupling corrections
to $S$ and $T$ imply a strong correlation between the fourth generation
mass scale and the largest allowed value for $|V_{t^{\prime }b}|\simeq s_{34}$.
The contour regions in the $S-T$ plane for a given confidence level $CL$ are
given by%
\begin{equation}
\chi ^{2}(S,T)=\frac{(S-S_{0})^{2}}{\sigma _{S}^{2}(1-\rho ^{2})}+\frac{%
(T-T_{0})^{2}}{\sigma _{T}^{2}(1-\rho ^{2})}+\frac{%
2(S_{0}T+ST_{0}-ST-S_{0}T_{0})}{\sigma _{S}\sigma _{T}(1-\rho ^{2})}<-2\ln(1-CL).  \label{8}
\end{equation}%

\bigskip

As an illustration, we compute the corresponding confidence levels for $S$
and $T$, by naively varying the quark masses ($m_{t^{\prime }}$,$m_{b^{\prime
}}$) within the range $[300,800]$ GeV. Only evaluated for the Higgs reference mass $m_{H}^{ref}=m_{Z}$, the LEPEWWG fit results $S_{0}\pm \sigma _{S}=0.05\pm 0.10\ $and $T_{0}\pm \sigma
_{T}=0.10\pm 0.09$ are used for the central values and their related
standard deviations ($\rho =0.85$ gives the correlation
coefficient between the two parameters) \cite{44}. The top reference mass is
defined as $m_{t}^{ref}=173.2$ GeV.\ All relevant quantities, including $%
\alpha ^{-1}(M_{Z})=128.89$ and $s_{W}^{2}(M_{Z})=0.22291$, are considered
at the Z pole. We have checked that our results are in agreement with \cite%
{42} at the permille level, yet very sensitive to the choice of $%
(S_{0},T_{0})$. As a result, the
upper limit on the sine of the mixing angle $\theta _{34}$ yields 
\begin{equation}
(s_{34})_{S,T}\leq \frac{m_{W}}{m_{t^{\prime }}},
\end{equation}
at the 95\% CL, in agreement with \cite{45}, though weaker than (\ref{10}).
The minimization of (\ref{8}) with respect to the $t^{\prime
} $ mass and the mixing angle is shown in Figure 2, as well as the upper bounds derived from $R_{b}
$.

\bigskip

Although they provide additional insight on the SM4 parameters, we stress that flavour physics constraints are not considered in this analysis. As we neglect the 1-4 and 2-4 CKM mixing, the electroweak precision observables are found to give more severe constraints on the maximally allowed values for 
$|V_{t^{\prime }b}|$. However, both the electroweak precision and the flavour constraints are known to favour small mixing with a heavy fourth family \cite{42}.

\begin{figure}[htbp]
\centering\includegraphics[scale=0.98]{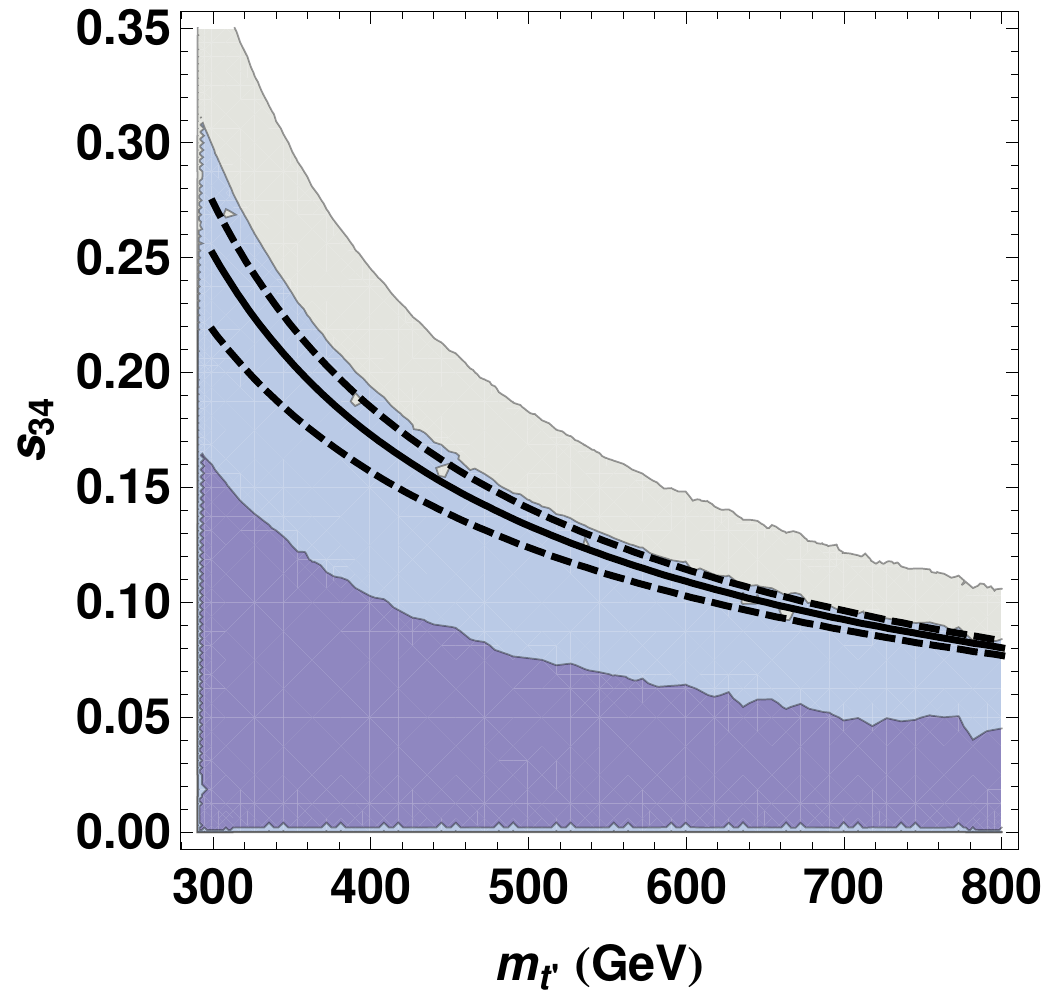}
\caption{68\%\ (darker blue), 95\%\ (blue) and 99\% (grey)
CL allowed regions from (\ref{8}) for the $3-4$ CKM mixing element as a function of $m_{t^{\prime }}$,
with all other SM4\ parameters unconstrained. We find that the mixing between the third and fourth
generation quarks is restricted such that $s_{34}<0.15$ $%
(0.075)$ holds at the 2$\protect\sigma $ (1$\protect\sigma )$ level if $%
m_{t^{\prime }}>500$ GeV when considering the $T$ parameter.\ The
non-oblique correction arising from the $Z\rightarrow b\bar{b}$ rate at the 95\%\ CL is
given by the solid black line for $\Delta m_{q^{\prime }}=0$ GeV.\ The upper and lower dashed lines give the
corresponding constraints for $\Delta m_{q^{\prime }}=-80$ GeV and $\Delta
m_{q^{\prime }}=80$ GeV, respectively.}
\label{}
\end{figure}

\newpage

\section{Parameter space}

In this section we present a systematic analysis of the SM4
parameter space, assessing the possibility that particular regions
might have evaded the experimental searches. Parameter sets minimizing the values of the electroweak parameters $S$ and $%
T $ are purposely identified, considering the relative importance of the
possible fermion mass hierarchies for the case of vanishing 2-4 and 1-4 family quark mixings. In this case, the mixing angle between the third and the fourth generation of quarks fully encodes the new flavour sector, and provides a simple baseline scenario to constrain efficiently the fourth
generation scenarios probed at the LHC. We require in particular $s_{34}<0.13$, given that such an upper bound is expected if the fourth generation quark masses are larger than 500 GeV. For much smaller mixings, the lightest fourth generation quark and lepton might be quasi-stable. Such long-lived fermions would then have distinct phenomenological signatures, and the current bounds should be re-examined \cite{46}. 

\newpage

As far as the scalar sector is concerned, the experimental bounds on the SM4 Higgs boson mass discussed in the previous section severely restrict the allowed parameter space. We select purposely two regions of interest for the Higgs
mass, namely%
\begin{eqnarray*}
\text{(a)\qquad }117\text{ GeV} &<&m_{H}<135\text{ GeV,%
} \\
\text{(b)\qquad }450\text{ GeV} &<&m_{H}<700\text{ GeV.%
}
\end{eqnarray*}
We show in Figure 3 the scatter plots of the $S$ and $T$ parameters, scanning the SM4 parameter space for these two scenarios. Assuming a normal quark mass hierarchy ($m_{t^{\prime
}}>m_{b^{\prime }}$), we consider $m_{t^{\prime
},b^{\prime }}\in \lbrack 350,700]$ GeV, $m_{\nu ^{\prime }}\in \lbrack
60,120]$ GeV and $m_{\tau ^{\prime }}\in \lbrack 100,250]$ GeV, and a light and a heavy Higgs separately.

\begin{figure}[htbp]
\centering\includegraphics[scale=0.37]{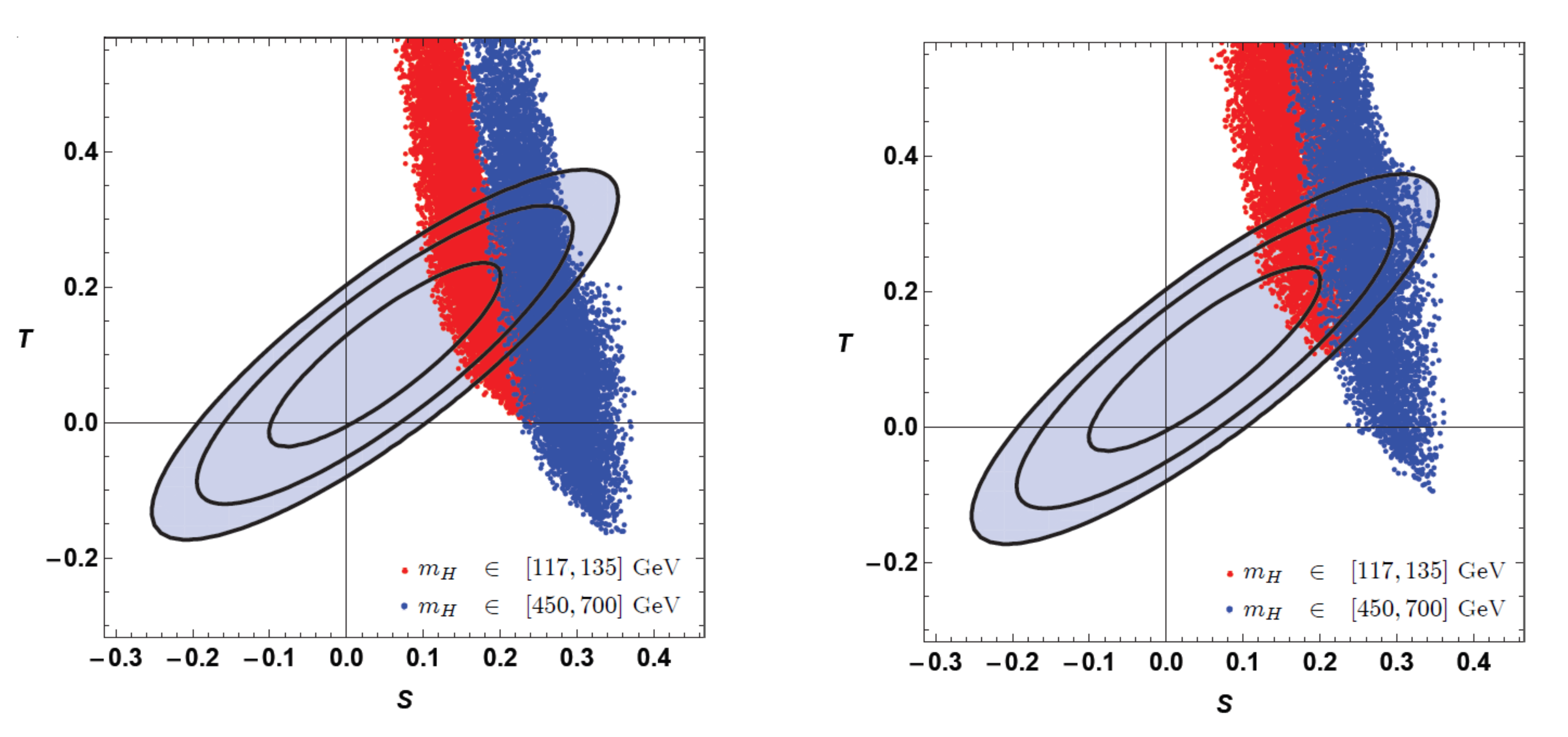}
\caption{$S$ and $T$ ellipses corresponding to $\Delta m_{q^{\prime }}<50$
GeV (left) and $\Delta m_{q^{\prime }}>50$ GeV (right) for the 68\%, 95\%
and the 99\% contour regions as defined by the inequality (\protect\ref{8}).\ Each
scatter plot depicts 10$^{6}$ random points for scenarios (a) in red and (b) in blue, under the constraints $350$ GeV$%
<m_{t^{\prime },b^{\prime }}<700$ GeV, $60$ GeV$<m_{\protect\nu ^{\prime
}}<120$ GeV, $100$ GeV$<m_{\protect\tau ^{\prime }}<250$ GeV and $%
0<s_{34}<0.13$. }
\label{Kribs}
\end{figure}

\bigskip

Favored by the electroweak precision fits, the case (a) is of first importance for the on-going searches at the LHC if the present excess in $H\rightarrow \gamma \gamma $ is a statistical fluctuation. On the other hand, (b) covers the allowed high Higgs mass region, whose upper limit $700$ GeV is chosen in agreement with the quantum triviality bound obtained for four chiral families, recently updated in \cite{12}. As our results indicate, none of the two quark mass differences $\Delta m_{q^{\prime
}}<50$ GeV and $\Delta m_{q^{\prime }}>50$ GeV is disfavored if a
small mixing $s_{34}$ is required. Although the
light Higgs case (a) is preferred at the 68\%\ confidence level, the scenario (b) is still
consistent at 95\%\ CL with a fourth generation of fermions. Both sets of points overlap
with a significant area of the confidence level ellipses, still allowing for
a large parameter space.\ The case $%
m_{t^{\prime }}<m_{b^{\prime }}$, not shown on the above picture, shifts the sets upwards by $\Delta
T\simeq 0.035$.\ 

\newpage

Regarding the Higgs mass dependence, the heavy case $m_{H}\simeq 650$ GeV is shifted towards $\Delta
S\simeq 0.08$ and $\Delta T\simeq -0.24$ with respect to the light
scenario $m_{H}\simeq 125$ GeV. Regarding the possible\ mass hierarchies, the
correlations between the fermion splittings $\Delta m_{q^{\prime
}}=m_{t^{\prime }}-m_{b^{\prime }}$ and $\Delta m_{l^{\prime }}=m_{\nu
^{\prime }}-m_{\tau ^{\prime }}$ have been considered in \cite{5}
for various quark and lepton mass scales. For completeness, we briefly re-examine them here for
a non-vanishing 3-4 CKM mixing. Fixing $m_{H}=[125,600]$ GeV, we plot in Figure\ 4\ the corresponding 68\%, 95\% and 99\% CL regions in the ($\Delta m_{q^{\prime }}$, $\Delta
m_{l^{\prime }}$) plane for all possible mass differences and CKM mixing elements.\ 

\bigskip \bigskip \bigskip

\begin{figure}[htbp]
\centering\includegraphics[scale=0.68]{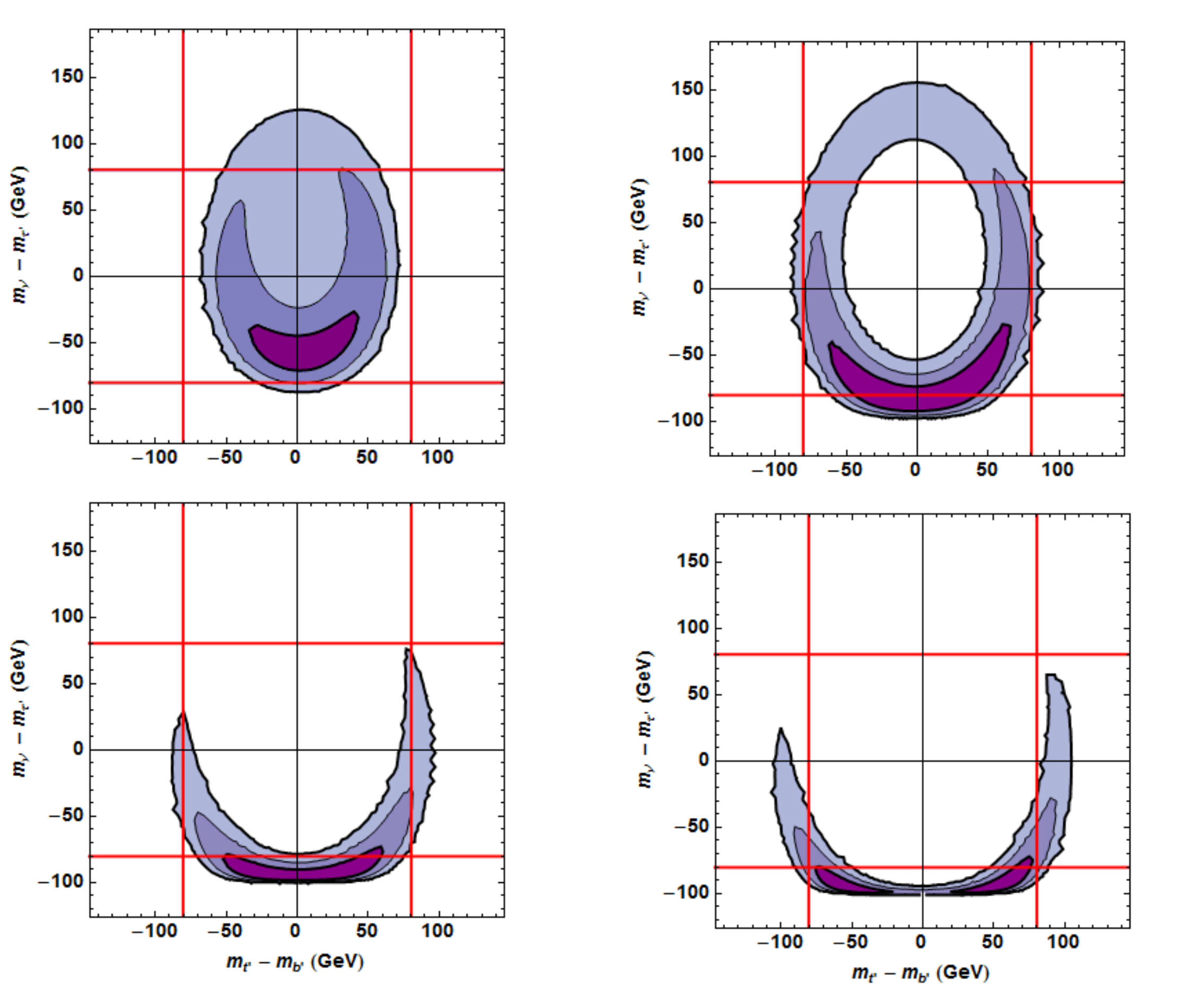}
\caption{Contour regions in the $\Delta m_{q^{\prime }}-\Delta m_{l^{\prime
}}$ plane as allowed from (\protect\ref{8}), for $m_{H}=125$ GeV (top), $600$
GeV (bottom), and $s_{34}=0.01$ (left), $0.1$ (right). The three confidence
levels 68\%, 95\% and 99\% are given by the purple, dark blue and light blue
areas, respectively, for $m_{t^{\prime }}=500\ $GeV and $m_{\protect\tau ^{\prime
}}=100\ $GeV. The vertical (horizontal) red lines define the $\Delta m_{q^{\prime }}=m_{W}$ $(\Delta m_{l^{\prime }}=m_{W})$ threshold for
the production of a real $W$ in $t^{\prime }\leftrightarrow b^{\prime }$ ($\tau ^{\prime }\leftrightarrow
\nu ^{\prime }$) decays.}
\label{}
\end{figure}
\newpage

Assuming a 500 (100) GeV scale for the fourth generation quarks (leptons), the
$\chi ^{2}$ values\ given by (\ref{8}) allow large
areas for $\Delta m_{q^{\prime }},\Delta m_{l^{\prime }}\geq 50$ GeV and $%
s_{34}\leq 10^{-1}$. Numerically, a strong correlation is confirmed between the quark and lepton
mass splittings, with a weak yet relevant dependence on the 3-4 CKM mixing. Given the logarithmic dependence of $S_{4}$ on the ratio $m_{\nu ^{\prime }}/m_{\tau ^{\prime }}$, large lepton mass splittings can be allowed at the cost of a smaller quark mass difference balancing the $T_{4}$ parameter.

\bigskip

We observe that the lepton mass hierarchy $m_{\nu ^{\prime}}>m_{\tau ^{\prime }}$ occurs to be strongly disfavored in view of these results,
whereas the quark splittings $\Delta m_{q^{\prime }}>0$ and $\Delta
m_{q^{\prime }}<0$ are equally preferred if $m_{\nu ^{\prime }}<m_{\tau
^{\prime }}$. Assuming that $s_{34}\geq 10^{-2}$, small lepton mass splittings are allowed in scenario (a) if $|\Delta m_{q^{\prime }}|$ is larger than 50 GeV, while $|\Delta m_{q^{\prime }}|<40$ GeV defines a wide parameter region for which $m_{\tau ^{\prime }}-m_{\nu ^{\prime }}>80$ GeV is possible. The oblique parameters thus allow the two-body decay mode $\tau ^{\prime }\rightarrow \nu ^{\prime }W$ for fourth generation leptons if the quark mass difference is small. However, the constraints obtained from Higgs searches severely restrict such a possibility in the low mass region.

\bigskip

Considering in particular the low mass region (a), we extract in Table 2 various representative benchmark points for $m_{H}=125$ GeV and the two cases $%
m_{t^{\prime }}>m_{b^{\prime }}$ and $m_{t^{\prime }}<m_{b^{\prime }}$. The heavy Higgs boson scenario (b) yields similar conclusions, though quark mass differences larger than $m_{W}$ are allowed at the 95\%\ confidence level inasmuch as the conditions $s_{34}<0.05$ and $m_{H}>400$ GeV are fulfilled simultaneously. Otherwise, quark mass splittings larger than 75 GeV are disfavored if $m_{H}<400$ GeV, which confirms the non trivial correlation between the Higgs mass and the quark mass difference depicted in Figure 1. 

\bigskip

Little change is noticed with respect to these conclusions when considering mixing angles smaller than $10^{-2}$. As Figure 1 indicates, there is not much gain in lowering down $s_{34}$ for the considered mass scales. We also observe that the upper bound on the quark mass splitting does not depend on any parameter other than the Higgs boson mass, unless the mixing angle saturates the upper bound required from $R_{b}$. If only the $T$ constraint is satisfied, both the quark and the lepton mass scales can be increased without any effect on the $S$ parameter. 

\bigskip \bigskip

\hspace{-1.0cm}
\begin{tabular}{|c|c|c|c|}
\hline
$m_{t\prime }$ & $\Delta m_{s_{34}=0.1}^{\max }$ & $\Delta
m_{s_{34}=0.05}^{\max }$ & $\Delta m_{s_{34}=0.01}^{\max }$ \\ \hline
400 & 62 & 68 & 70 \\ \hline
500 & 52 & 66 & 70 \\ \hline
600 & 36 & 64 & 70 \\ \hline
\end{tabular}
\begin{tabular}{|c|c|c|c|}
\hline
$m_{b\prime }$ & $\Delta m_{s_{34}=0.1}^{\max }$ & $\Delta
m_{s_{34}=0.05}^{\max }$ & $\Delta m_{s_{34}=0.01}^{\max }$ \\ \hline
400 & -65 & -72 & -72 \\ \hline
500 & -55 & -70 & -72 \\ \hline
600 & -38 & -66 & -72 \\ \hline
\end{tabular}

\bigskip 

$\ \qquad \qquad \qquad \qquad m_{t^{\prime }}>m_{b^{\prime }}$ \ \qquad
\qquad \qquad \qquad \qquad \qquad\ $m_{t^{\prime }}<m_{b^{\prime }}$

\bigskip 

Table 2\ : Selected SM4\ scenarios, allowed at the 95\%\ CL\ by the $S$ and $T$ parameters, for $m_{H}=125$ GeV, and $s_{34}=0.1,0.05,0.01$.  The shown benchmark points give the largest possible quark mass splittings $\Delta m_{q^{\prime
}}^{\max }=m_{t^{\prime }}-m_{b^{\prime }}$, considering
both the normal ($m_{t^{\prime }}>m_{b^{\prime }}$) and the inverted ($%
m_{t^{\prime }}<m_{b^{\prime }}$) mass hierarchies. All masses are given in GeV, with $m_{\nu
^{\prime }}=60$ GeV and $\Delta m_{l^{\prime }}=40$ GeV.

\bigskip

\section{Phenomenology at the LHC}

The predominant production and decay modes for a fourth generation of quarks strongly depend on the quark mass splitting and hierarchy. Assuming a small 3-4 CKM mixing, we here consider the corresponding production modes, decay channels, and LHC signatures.

\subsection{Production}

Although pair production via the strong interaction is the main mechanism for producing new heavy quarks at the LHC, the electroweak production of\ a single top quark in the $t$-channel provides another relevant process to search for a fourth quark family \cite{47}. For instance, the
final state topologies expected from the electroweak $t^{\prime }\bar{b}$ ($%
\bar{t}^{\prime }b$) and $\bar{t}b^{\prime }$ ($t\bar{b}^{\prime }$)
production have been used recently in \cite{10} to set very restrictive
bounds on a degenerate fourth generation. However, these processes scale by a factor $|V_{t^{\prime }b}|^{2}\simeq
|V_{tb^{\prime }}|^{2}$ and are strongly suppressed if the 3-4 CKM mixing element is
negligibly small (while pair production is independent of it). Considering
small $s_{34}$ values as discussed in the previous section, single $%
t^{\prime }$ and $b^{\prime }$ production suffer strong CKM suppression,
with cross sections below the femtobarn level for the considered mass scales
and mixings. Yet, the $t^{\prime }\bar{b}^{\prime }$ ($\bar{t}^{\prime }b^{\prime }$)
electroweak electroweak production in the $t$-channel provides
an alternative without any CKM\ suppression, which overtakes $t^{\prime }b$ ($tb^{\prime }$) single
production by two orders of magnitude if $s_{34}$ is smaller
than $10^{-1}$.

\begin{figure}[htbp]
\centering\includegraphics[scale=0.15]{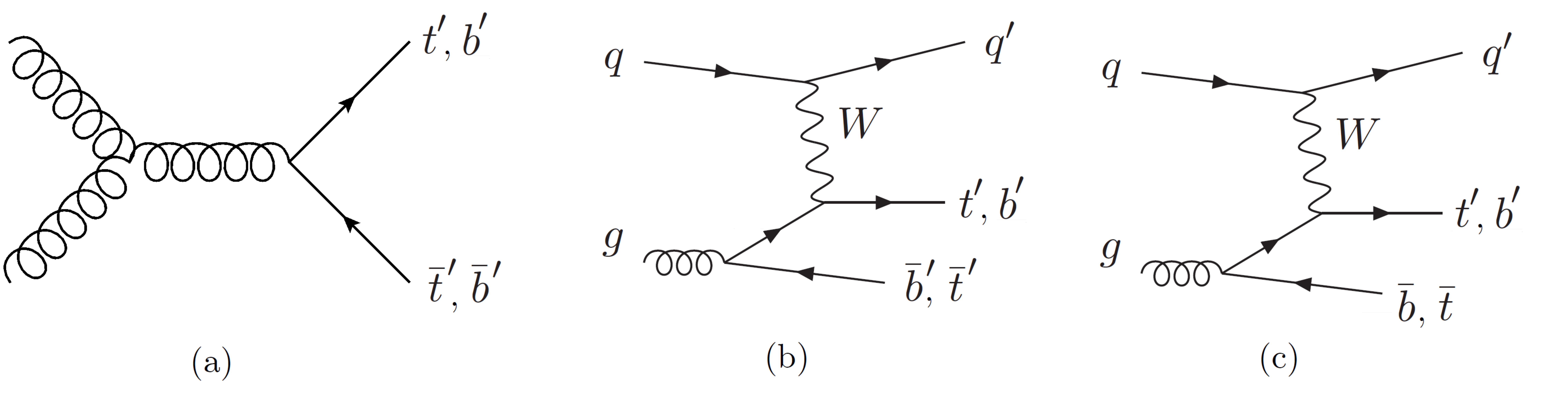}
\caption{Representative diagrams for the production of fourth
generation quarks. (a) Strong pair production of heavy quarks at leading order ; (b) $2\rightarrow 3$ Born diagrams contributing to $t^{\prime
}\bar{b}^{\prime }$ and $\bar{t}^{\prime }b^{\prime }$ electroweak
production in the $t$-channel ; (c) $t^{\prime
}\bar{b}$ and $\bar{t}b^{\prime }$ electroweak production (and the charge-conjugate processes) in the $t$-channel (suppressed if a small 3-4 CKM mixing is considered).}
\label{}
\end{figure}

Note further that two heavy particles are present in the final state, the interaction
with the longitudinal $W$ boson is proportional to the $t^{\prime }$ and $b^{\prime }$
Yukawa couplings and dominates the electroweak production of heavy
quarks.\ This channel features the second dominant signal after
strong pair production when the fourth generation mixings with the lighter
quarks are negligible. Benchmark cross-sections at NLO\ have been tabulated
in \cite{47} for various fourth generation masses, PDF sets and LHC\
energies under the assumption $|V_{t^{\prime }b^{\prime }}|^{2}=1$. Tables 3, 4 and 5
present our predictions for relevant parameter choices.\ The $%
t^{\prime }\bar{b}^{\prime }$ ($\bar{t}^{\prime }b^{\prime }$) $t$-channel
electroweak production, and the pair production cross-sections have been computed at 7, 8 and 14 TeV, with the CTEQ6.6 PDF set. The 6.0\ version of the Monte-Carlo
software MCFM \cite{48} has been used for the calculations at NLO accuracy,
following the conventions used previously in \cite{47}.

\bigskip

\begin{equation*}
\begin{tabular}{|c|c|c|c|c|}
\hline
$m_{t^{\prime }}$ & $m_{b^{\prime }}$ & $\sigma _{t^{\prime }\bar{b}^{\prime
}}^{7\text{ TeV}}(fb)$ & $\sigma _{t^{\prime }\bar{b}^{\prime }}^{8\text{ TeV%
}}(fb)$ & $\sigma _{t^{\prime }\bar{b}^{\prime }}^{14\text{ TeV}}(fb)$ \\ 
\hline
400 & 350 & (100) 120$_{-6\text{ }-10}^{+12\text{ }+13}$ & (155) 262$_{-6%
\text{ }-16}^{+14\text{ }+18}$ & (1162) 1395$_{-28\text{ }-61}^{+55\text{ }%
+69}$ \\ \hline
400 & 330 & (111) 188$_{-9\text{ }-15}^{+18\text{ }+21}$ & (187) 223$_{-13%
\text{ }-20}^{+22\text{ }+26}$ & (1151) 1512$_{-31\text{ }-64}^{+48\text{ }%
+66}$ \\ \hline
400 & 310 & (121) 203$_{-7\text{ }-14}^{+15\text{ }+17}$ & (201) 265$_{-18%
\text{ }-25}^{+29\text{ }+33}$ & (1218) 1720$_{-36\text{ }-77}^{+60\text{ }%
+81}$ \\ \hline
&  &  &  &  \\ \hline
500 & 450 & (32.9) 49.7$_{-4.6\text{ }-6.1}^{+7.8\text{ }+8.8}$ & (57.1) 92.6%
$_{-6.4\text{ }-9.3}^{+11.1\text{ }+12.6}$ & (468) 610$_{-17\text{ }-35}^{+32%
\text{ }+40}$ \\ \hline
500 & 430 & (34.8) 50.1$_{-5.0\text{ }-6.5}^{+8.7\text{ }+9.7}$ & (67.2) 87.2%
$_{-6.2\text{ }-8.6}^{+11.0\text{ }+12.4}$ & (535) 653$_{-19\text{ }-37}^{+34%
\text{ }+42}$ \\ \hline
500 & 410 & (40.9) 48.7$_{-4.5\text{ }-6.0}^{+7.6\text{ }+8.5}$ & (70.9) 113$%
_{-7\text{ }-10}^{+12\text{ }+14}$ & (614) 719$_{-16\text{ }-37}^{+32\text{ }%
+41}$ \\ \hline
&  &  &  &  \\ \hline
600 & 550 & (11.7) 19.1$_{-2.2\text{ }-2.9}^{4.1\text{ }+4.6}$ & (24.2) 33.6$%
_{-3.5\text{ }-4.5}^{+5.8\text{ }+6.5}$ & (278) 390$_{-9\text{ }-22}^{+21%
\text{ }+25}$ \\ \hline
600 & 530 & (12.4) 16.7$_{-2.1\text{ }-2.7}^{+3.8\text{ }+4.2}$ & (26.1) 38.0%
$_{-4.3\text{ }-5.5}^{+7.8\text{ }+8.6}$ & (301) 367$_{-9\text{ }-21}^{+19%
\text{ }+23}$ \\ \hline
600 & 510 & (14.1) 18.7$_{-2.4\text{ }-3.0}^{+4.2\text{ }+4.6}$ & (28.2) 36.2%
$_{-3.4\text{ }-4.6}^{+6.3\text{ }+6.9}$ & (321) 323$_{-17\text{ }-26}^{+30%
\text{ }+35}$ \\ \hline
\end{tabular}%
\end{equation*}

\bigskip

Table 3: $\sqrt{s}=\ 7,$\ $8$ and $14$ TeV NLO predictions for the $t^{\prime }\bar{b}^{\prime }$ t-channel electroweak production, assuming $|V_{t^{\prime }b^{\prime }}|^{2}=1$. LO predictions are given in parentheses. The first
uncertainty is calculated from the renormalisation and factorisation scales
variation using the conventions detailed in \cite{47}, while the second gives the
upper and lower PDF errors for the CTEQ6.6 PDF set. The central value is computed for $\mu =\mu _{f}=\mu _{r}=(m_{t^{\prime }}+m_{b^{\prime }})/2$. The assumed CP\
invariance and chirality of the $Wt^{\prime }b^{\prime }$ vertex lead to
equal cross-sections when switching the $t^{\prime }$ and the $%
b^{\prime }$ masses, here given in GeV.

\bigskip \bigskip \bigskip

\begin{equation*}
\begin{tabular}{|c|c|c|c|c|}
\hline
$m_{t^{\prime }}$ & $m_{b^{\prime }}$ & $\sigma _{\bar{t}^{\prime }b^{\prime
}}^{7\text{ TeV}}(fb)$ & $\sigma _{\bar{t}^{\prime }b^{\prime }}^{8\text{ TeV%
}}(fb)$ & $\sigma _{\bar{t}^{\prime }b^{\prime }}^{14\text{ TeV}}(fb)$ \\ 
\hline
400 & 350 & (39.7) 52.7$_{-2.5\text{ }-4.6}^{+7.0\text{ }+9.4}$ & (75.2) 102$%
_{-3\text{ }-7}^{+11\text{ }+14.6}$ & (537) 793$_{-18\text{ }-36}^{+46\text{ 
}+68}$ \\ \hline
400 & 330 & (50.4) 78.0$_{-4.1\text{ }-6.4}^{+10.4\text{ }+13.3}$ & (84.8)
109$_{-5\text{ }-9}^{+14\text{ }+18}$ & (587) 602$_{-16\text{ }-32}^{+40%
\text{ }+57}$ \\ \hline
400 & 310 & (52.0) 64.7$_{-4.0\text{ }-6.2}^{+9.4\text{ }+12.0}$ & (87.7) 112%
$_{-6\text{ }-10}^{+14\text{ }+18}$ & (676) 934$_{-20\text{ }-44}^{+45\text{ 
}+77}$ \\ \hline
&  &  &  &  \\ \hline
500 & 450 & (12.2) 18.9$_{-1.7\text{ }-2.4}^{+4.0\text{ }+5.0}$ & (23.4) 42.3%
$_{-1.6\text{ }-3.8}^{+5.9\text{ }+8.1}$ & (239) 300$_{-9\text{ }-17}^{+23%
\text{ }+32}$ \\ \hline
500 & 430 & (13.9) 21.8$_{-2.0\text{ }-2.7}^{+4.5\text{ }+5.5}$ & (26.2) 44.2%
$_{-2.4\text{ }-3.8}^{+6.2\text{ }+8.0}$ & (287) 257$_{-8\text{ }-16}^{+21%
\text{ }+29}$ \\ \hline
500 & 410 & (15.9) 48.6$_{-2.5\text{ }-4.1}^{+6.3\text{ }+8.3}$ & (31.2) 48.6%
$_{-2.5\text{ }-4.1}^{+6.3\text{ }+8.3}$ & (293) 327$_{-9\text{ }-19}^{+25%
\text{ }+35}$ \\ \hline
&  &  &  &  \\ \hline
600 & 550 & (4.07) 6.68$_{-0.81\text{ }-1.10}^{+1.85\text{ }+2.26}$ & (9.5)
12.8$_{-1.2\text{ }-1.6}^{+2.7\text{ }+3.3}$ & (117) 149$_{-5\text{ }%
-10}^{+15\text{ }+20}$ \\ \hline
600 & 530 & (4.29) 6.72$_{-0.73\text{ }-0.97}^{+1.74\text{ }+2.08}$ & (9.8)
11.8$_{-1.2\text{ }-1.6}^{+2.7\text{ }+3.3}$ & (121) 142$_{-6\text{ }%
-10}^{+14\text{ }+19}$ \\ \hline
600 & 510 & (5.23) 7.83$_{-0.86\text{ }-1.11}^{+1.99\text{ }+2.36}$ & (10.9)
15.0$_{-1.3\text{ }-1.9}^{+3.2\text{ }+3.9}$ & (142) 159$_{-6\text{ }%
-12}^{+17\text{ }+23}$ \\ \hline
\end{tabular}%
\end{equation*}

Table 4: Same as Table 3 for $t$-channel $\bar{t}%
^{\prime }b^{\prime }$ electroweak production, with the CTEQ6.6 PDF set and $|V_{t^{\prime }b^{\prime }}|^{2}=1$. 

\bigskip

\begin{equation*}
\begin{tabular}{|c|c|c|c|}
\hline
$m_{Q}$ & $\sigma _{Q\bar{Q}}^{7\text{ TeV}}(fb)$
& $\sigma _{Q\bar{Q}}^{8\text{ TeV}}(fb)$ & $\sigma
_{Q\bar{Q}}^{14\text{ TeV}}(fb)$ \\ \hline
400 & (955) 1329$_{-138\text{ }-179}^{+240\text{ }+280}$ & (1552) 2164$_{-202%
\text{ }-267}^{+341\text{ }+400}$ & (9206) 13033$_{-650\text{ }-1028}^{+1063%
\text{ }-1169}$ \\ \hline
425 & (654) 907$_{-98\text{ }-127}^{+175\text{ }+202}$ & (1078) 1498$_{-147%
\text{ }-192}^{+251\text{ }+293}$ & (6675) 9414$_{-511\text{ }-784}^{+836%
\text{ }+1038}$ \\ \hline
450 & (454) 629$_{-71\text{ }-91}^{+127\text{ }+148}$ & (761) 1054$_{-108%
\text{ }-140}^{+188\text{ }+218}$ & (4913) 6905$_{-404\text{ }-604}^{+661%
\text{ }+812}$ \\ \hline
475 & (320) 442$_{-51\text{ }-65}^{+95\text{ }+109}$ & (544) 751$_{-80\text{ 
}-103}^{+141\text{ }+164}$ & (3665) 5131$_{-321\text{ }-469}^{+526\text{ }%
+641}$ \\ \hline
500 & (228) 314$_{-37.1\text{ }-48}^{+70.8\text{ }+81}$ & (393) 542$_{-59%
\text{ }-76}^{+107\text{ }+124}$ & (2767) 3861$_{-257\text{ }-368}^{+422%
\text{ }+510}$ \\ \hline
525 & (164) 226$_{-27\text{ }-35}^{+53\text{ }+61}$ & (287) 395$_{-44\text{ }%
-57}^{+82\text{ }+94}$ & (2112) 2938$_{-207\text{ }-291}^{+340\text{ }+408}$
\\ \hline
550 & (119) 164$_{-20\text{ }-26}^{+41\text{ }+46}$ & (212) 291$_{-34\text{ }%
-43}^{+63\text{ }+73}$ & (1628) 2258$_{-167\text{ }-232}^{+276\text{ }+329}$
\\ \hline
575 & (87.2) 120$_{-15\text{ }-19}^{+31\text{ }+35}$ & (158) 216$_{-25\text{ 
}-33}^{+49\text{ }+56}$ & (1266) 1750$_{-135\text{ }-198}^{+225\text{ }+267}$
\\ \hline
600 & (64.4) 88.4$_{-11\text{ }-14}^{+24\text{ }+27}$ & (119) 162$_{-19\text{ 
}-25}^{+38\text{ }+44}$ & (993) 1369$_{-110\text{ }-149}^{+184\text{ }+218}$
\\ \hline
\end{tabular}%
\end{equation*}%
Table 5: (LO) NLO cross-sections for $%
Q=t^{\prime },b^{\prime }$ pair production at the LHC\ 7, 8 and 14
TeV. The first error gives the factorization and renormalization
scales dependence, computed by setting $\mu =\mu _{f}=\mu _{r}\ $and varying
them between $m_{Q}/2$ and $2m_{Q}\ $for the upper and
lower deviations from the central value $\mu =m_{Q}$, respectively.
The second error yields the CTEQ6.6 PDF uncertainties.

\bigskip

\subsection{Decay}

Large quark mass differences open new possibilities in the
aforementioned benchmark scenarios. Even if $|m_{t^{\prime }}-m_{b^{\prime
}}|<m_{W}$, the three-body $t^{\prime }\rightarrow b^{\prime }W^{(\ast )}$ ($%
b^{\prime }\rightarrow t^{\prime }W^{(\ast )}$) decay can overcome the
two-body $qW$ modes at tree-level if the magnitude of the CKM\ entry $%
V_{t^{\prime }q}$ $(V_{qb^{\prime }})$ at which the $t^{\prime }\rightarrow
Wq$ ($b^{\prime }\rightarrow Wq$) decay proceeds is small enough. If
the fourth generation quarks are allowed to decay into each other, the heavy-to-heavy
transition $t^{\prime }\rightarrow b^{\prime }W$ ($b^{\prime }\rightarrow
t^{\prime }W$) competes with the CKM\ suppressed decay $t^{\prime
}\rightarrow bW$ ($b^{\prime }\rightarrow tW$) for $|V_{t^{\prime }b}|\simeq
|V_{tb^{\prime }}|\ll |V_{t^{\prime }b^{\prime }}|$ \cite{49}. Assuming
that all off-shell particles are stable with a negligible width, the
two-body decay width of a heavy quark $t^{\prime }$ is given by
\begin{eqnarray}
\Gamma (t^{\prime } &\rightarrow &Wq)=\frac{G_{F}\text{ }m_{t^{\prime }}^{3}%
}{8\pi \sqrt{2}}|V_{t^{\prime }q}|^{2}\text{ }f_{2}\Big(\frac{m_{q}}{%
m_{t^{\prime }}},\frac{m_{W}}{m_{t^{\prime }}}\Big),  \label{25} \\
f_{2}(\alpha ,\beta ) &=&[(1-\alpha ^{2})^{2}+\beta ^{2}(1+\alpha
^{2})-2\beta ^{4}]\sqrt{[1-(\alpha +\beta )^{2}][1-(\alpha -\beta )^{2}].} 
\notag
\end{eqnarray}%
If $m_{t^{\prime }}-m_{b^{\prime }}<m_{W\text{ }}$ (which defines the real $W$
production threshold), $t^{\prime }\rightarrow b^{\prime }W^{(\ast )}$ ($%
b^{\prime }\rightarrow t^{\prime }W^{(\ast )}$) is phase-space suppressed, the partial rate (\ref{25}) is not
justified. Following \cite{49}%
, we require the $t^{\prime }$ $(b^{\prime })$ quarks to be on-shell particles in the
final state for the heavy-to-heavy transitions. For a
normal quark mass hierarchy ($m_{t^{\prime }}>m_{b^{\prime }}$), the formula
accounting for virtual and real $W$\ emission ($\Gamma _{W}\neq 0$ with $%
b^{\prime }$ assumed on-shell) reads \cite{46}
\begin{eqnarray}
\Gamma (t^{\prime } &\rightarrow &b^{\prime }W^{(\ast )})=\frac{%
G_{F}^{2}m_{t^{\prime }}^{5}}{192\pi ^{3}}|V_{t^{\prime }b^{\prime }}|^{2}%
\text{ }f\Big(\frac{m_{t^{\prime }}^{2}}{m_{W}^{2}},\frac{m_{b^{\prime }}^{2}}{%
m_{W}^{2}},\frac{\Gamma _{W}^{2}}{m_{W}^{2}}\Big),  \label{9} \\
f(\alpha ,\beta ,\gamma ) &=&2\int_{0}^{(1-\sqrt{\beta })^{2}}dx\text{ }%
\frac{[(1-\beta )^{2}+x\text{ }(1+\beta )-2x^{2}]\sqrt{\lambda (1,x,\beta )}%
}{[(1-\alpha x)^{2}+\gamma ^{2}]},  \notag
\end{eqnarray}
where $\lambda (a,b,c)=a^{2}+b^{2}+c^{2}-2ab-2bc-2ac$. The case
for $m_{b^{\prime }}>m_{t^{\prime }}$ is similar.

\newpage

We depict in Figure\ 6 the relative fraction $\Gamma
_{t^{\prime }\rightarrow b^{\prime }W^{(\ast )}}/\Gamma _{t^{\prime
}\rightarrow bW}$ versus the mass difference $\Delta m_{q^{\prime }}$ for a
normal quark mass ordering. The width effect occuring in the
decay of a virtual $W$ boson resonance near threshold is clearly visible for 
$m_{t^{\prime }}\approx m_{W}+m_{b^{\prime }}$. Table 6 displays the
decay rates and relative ratios obtained from (\ref{25}) and (\ref{9}), considering $s_{34}=0.1$ and $0.05$. Under these assumptions, the decays $t^{\prime }\rightarrow b^{\prime }W$ and $%
b^{\prime }\rightarrow t^{\prime }W$ can compete with the $4-3 $ transitions for
quark mass differences larger than $m_{W}$, while the heavy-to-heavy transitions
systematically dominate if $s_{34}<10^{-3}$ and $\Delta m_{q^{\prime }}>40$ GeV. The relative fraction $\Gamma _{b^{\prime }\rightarrow t^{\prime }W^{(\ast
)}}/\Gamma _{b^{\prime }\rightarrow tW}$ versus $m_{b^{\prime
}}-m_{t^{\prime }}$ leads to similar results when considering the same mass scale.

\bigskip
\begin{figure}[htbp]
\hspace{-1.0cm}
\centering\includegraphics[scale=0.55]{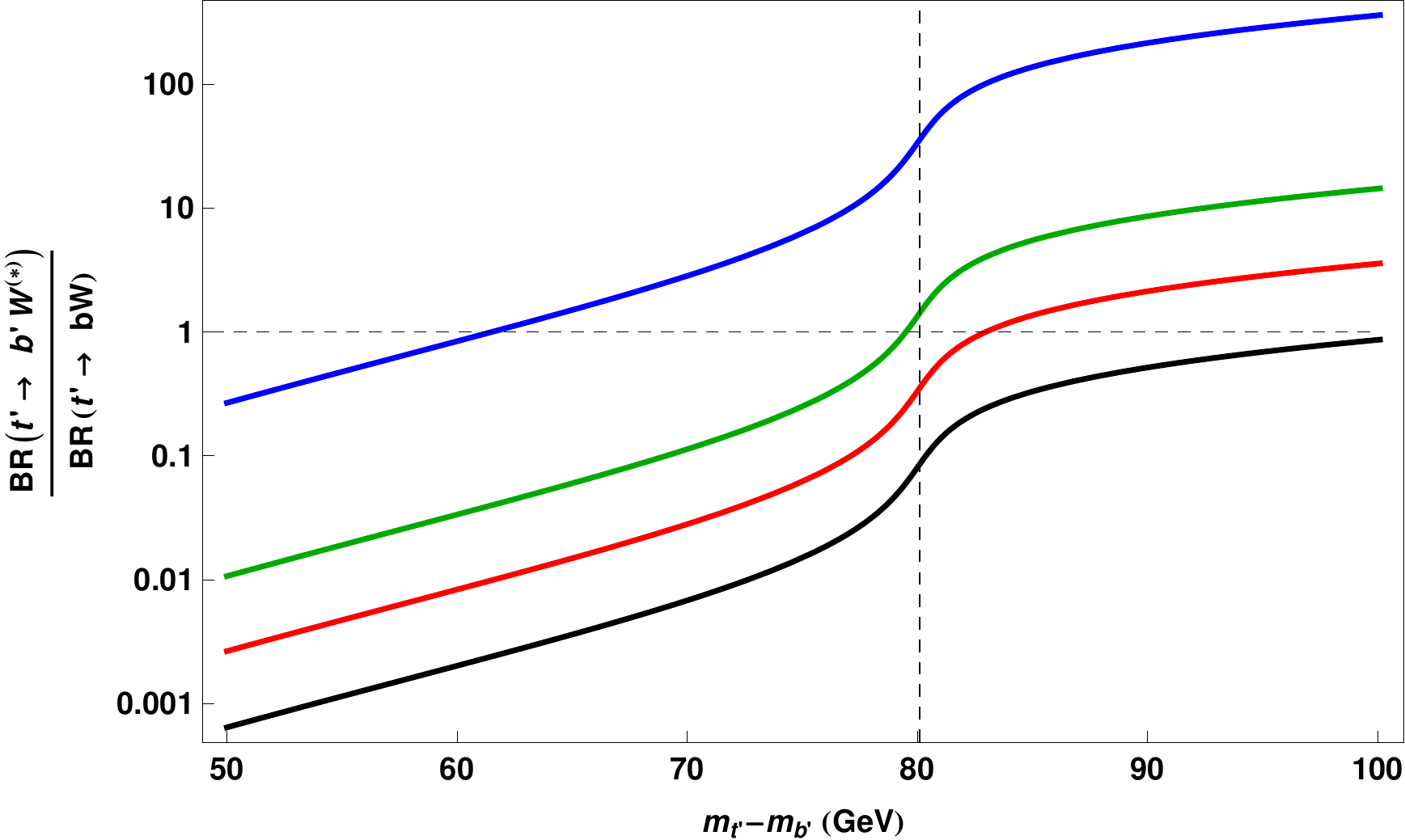}
\caption{Ratio of the $t^{\prime }\rightarrow
b^{\prime }W^{(\ast )}$ and $t^{\prime }\rightarrow bW$ decay widths for a
classical quark mass ordering.\ The case $m_{t^{\prime }}=500$ GeV is
displayed, with $|V_{t^{\prime }b}|=0.2$ (black), $0.1$ (red), $0.05$
(green) and $0.01$ (blue). }
\label{}
\end{figure}

\begin{equation*}
\begin{tabular}{|c|c|c|c|c|}
\hline
$m_{t^{\prime }}$ & $m_{b^{\prime }}$ & $\Gamma _{t^{\prime }\rightarrow bW}$
& $\Gamma _{t^{\prime }\rightarrow b^{\prime }W^{(\ast )}}$ & BR$(t^{\prime
}\rightarrow b^{\prime }W^{(\ast )})/$BR$(t^{\prime }\rightarrow bW)$ \\ 
\hline
400 & 350 & 0.203 (0.0507) & 0.00101 (0.00102) & 0.00499 (0.0201) \\ \hline
400 & 330 & " & 0.0105 (0.0106) & 0.0519 (0.209) \\ \hline
400 & 310 & " & 0.781 (0.787) & 3.850 (15.5) \\ \hline
&  &  &  &  \\ \hline
500 & 450 & 0.398 (0.0995) & 0.00105 (0.00106) & 0.00265 (0.0107) \\ \hline
500 & 430 & " & 0.0112 (0.0113) & 0.0281 (0.113) \\ \hline
500 & 410 & " & 0.848 (0.854) & 2.13 (8.59) \\ \hline
&  &  &  &  \\ \hline
600 & 550 & 0.688 (0.172) & 0.00108 (0.00109) & 0.00157 (0.00634) \\ \hline
600 & 530 & " & 0.0116 (0.0117) & 0.0169 (0.0680) \\ \hline
600 & 510 & " & 0.894 (0.900) & 1.30 (5.23) \\ \hline
\end{tabular}%
\end{equation*}

Table 6 : $t^{\prime }\rightarrow bW$ and $t^{\prime }\rightarrow b^{\prime
}W^{(\ast )}$ decay widths in GeV, and their relative ratio for $%
m_{t^{\prime }}>m_{b^{\prime }}$ for various
splitting scenarios and $s_{34}=0.1$ ($0.05$). 

\bigskip 

\subsection{Signatures at the LHC}

We now briefly discuss the possible decay signatures depending on
their mass ordering for small $s_{34}$ mixings.\ In the case of a normal quark mass hierarchy ($m_{t^{\prime }}>m_{b^{\prime }}$), up-like fourth
generation quarks either decay via $t^{\prime }\rightarrow bW$ or
cascade through $t^{\prime }\rightarrow b^{\prime }W\rightarrow
tWW\rightarrow bWWW$, with a real or a virtual $W$ emission. The
respective CKM\ mixing matrix factors $|V_{t^{\prime }b}|^{2}$ and $%
|V_{t^{\prime }b^{\prime }}|^{2}|V_{tb^{\prime }}|^{2}|V_{tb}|^{2}$ being of
the same magnitude, the $t^{\prime
}\rightarrow b^{\prime }W^{\ast }$ decay
can dominate over the $t^{\prime }\rightarrow Wb$ mode, though $\Gamma _{t^{\prime }\rightarrow b^{\prime
}W^{(\ast )}}$ is strongly phase space\ suppressed \cite{49}. If we enforce that BR$(t^{\prime }\rightarrow b^{\prime }W)\gg
$ BR$(t^{\prime }\rightarrow bW)$ and postulate that $t^{\prime }$ and $b^{\prime }$ decay promptly within the detector, the most likely final states for a normal mass hierarchy are

\begin{itemize}
\item (a) $t^{\prime }\bar{t}^{\prime }\rightarrow b^{\prime }\bar{b}%
^{\prime }W^{+}W^{-}\rightarrow t\bar{t}$ $2W^{+}2W^{-}\rightarrow b\bar{b}%
3W^{+}3W^{-},$

\item (b) $b^{\prime }\bar{b}^{\prime }\rightarrow t\bar{t}$ $%
W^{+}W^{-}\rightarrow b\bar{b}2W^{+}2W^{-},$

\item (c) $t^{\prime }\bar{b}^{\prime }\rightarrow b^{\prime }\bar{t}%
W^{+}W^{-}\rightarrow t\bar{b}2W^{+}2W^{-}\rightarrow b\bar{b}3W^{+}2W^{-},$

\item (d) $\bar{t}^{\prime }b^{\prime }\rightarrow \bar{b}^{\prime
}tW^{-}W^{+}\rightarrow \bar{t}b2W^{-}2W^{+}\rightarrow b\bar{b}%
3W^{-}2W^{+}. $
\end{itemize}

\bigskip

Table 7 includes the branching fractions corresponding to these various signatures. Each of the above final states includes two $b$-jets
for singly and pair produced fourth generation quarks, while four to
six W bosons occur in the decay of each signal event. These topologies all allow for same-sign ($l^{\pm
}l^{\pm }$) and opposite-sign ($l^{\pm }l^{\mp }$) dileptons in the final
state, with substantial missing energy along the jets. Interestingly, $t^{\prime }$ pair production (a) provides a signal with up to 12 jets if all $W$ bosons decay hadronically. Looking for the rare signature provided by two same-sign leptons with
large missing momentum is then a relevant search strategy, though a small
number of $l^{\pm }l^{\pm }$ events is expected. Except for $b^{\prime
}\ $pair production (b), we observe that the $%
t^{\prime }$ can be reconstructed from same-sign dileptonic events, as
at least two same-charge W's follow from the decay chain $t^{\prime
}\rightarrow b^{\prime }W\rightarrow tWW\rightarrow bWWW$.\ 

\bigskip

We also note that none of the
above channels include same-charge $W$ bosons decaying from the same $%
b^{\prime }$. The corresponding final states include four $W$ bosons
and two $b$ quarks. For $m_{t^{\prime }}>m_{b^{\prime }}$, the $b^{\prime }$
quark is not allowed to decay to $t^{\prime }W$ and the CKM\ suppressed $%
b^{\prime }\rightarrow tW$ becomes the only possible decay mode. However, the requirement
of two like-sign dilepton pairs provides this channel with a rare SM
signature containing very few background events. Searching for a
single lepton and trileptons is also a possible strategy. 

Although the relative cross section is expected to be smaller than pair production, $t^{\prime }b^{\prime }$ electroweak production (c,d) features
an interesting situation with up to 10 jets +\ 2 b-tags, for which an odd
number of same-sign leptons can be expected, two of them decaying from the same $t^{\prime }$ quark. Although appropriate cuts can improve the discrimination, the overwhelming combinatorial background makes the reconstruction and selection challenging issues. 

\newpage

An inverted mass hierarchy ($%
m_{t^{\prime }}<m_{b^{\prime }}$) would feature a down-type quark decaying either
as $b^{\prime }\rightarrow t^{\prime }W\rightarrow bWW$ or $b^{\prime
}\rightarrow tW\rightarrow bWW$, leading to identical final states. The two
corresponding decay rates are suppressed by $|V_{t^{\prime }b^{\prime
}}|^{2}|V_{t^{\prime }b}|^{2}$ and $|V_{tb^{\prime }}|^{2}|V_{tb}|^{2}$
respectively, both proportional to $s_{34}^{2}$. Assuming BR$(b^{\prime }\rightarrow
t^{\prime }W)\gg $ BR$(b^{\prime }\rightarrow tW)$, two to four $W$ bosons
can be present in the possible final states

\begin{itemize}
\item (e) $t^{\prime }\bar{t}^{\prime }\rightarrow b\bar{b}W^{+}W^{-}$

\item (f) $b^{\prime }\bar{b}^{\prime }\rightarrow t^{\prime }\bar{t}%
^{\prime }$ $W^{-}W^{+}\rightarrow b\bar{b}2W^{-}2W^{+}$

\item (g) $t^{\prime }\bar{b}^{\prime }\rightarrow b\bar{t}^{\prime
}W^{+}W^{-}\rightarrow b\bar{b}W^{+}2W^{-}$

\item (h) $\bar{t}^{\prime }b^{\prime }\rightarrow \bar{b}t^{\prime
}W^{-}W^{+}\rightarrow b\bar{b}2W^{+}W^{-}$
\end{itemize}

We observe that both the $t^{\prime }\bar{t}^{\prime }$ and the $b^{\prime }\bar{b}%
^{\prime }$ pair production signals (e,f) are covered by the direct searches \cite{8,9,10,11}
in multiple channels under the
condition BR$(t^{\prime }\rightarrow bW)=$ BR$(b^{\prime }\rightarrow tW)=100\%$%
. If a small $s_{34}^{2}$ value is assumed, the above channels final states are
however overwhelmed by the SM background, weakening the current bounds on $m_{t^{\prime }}$ and $m_{b^{\prime }}$. The $t^{\prime }b^{\prime }$ production processes (g,h) allow for 3 $%
W $ bosons and 2 b-tags in the final state. Despite the fact that the associated
cross-sections are low and hardly competitive with pair production, the
high-jet multiplicity provides again an interesting signature with the
possibility of single leptons, same-sign dileptons, and trileptons in the final state.

\bigskip \bigskip \bigskip \bigskip
\hspace{1.0cm}
\begin{tabular}{|c|c|c|c|c|c|c|c|}
\hline
& $6l$ & $5l$ & $4l$ & $3l$ & $2l$ & $1l$ & $0l$ \\ \hline
BR($t^{\prime }\bar{t}^{\prime }\rightarrow b\bar{b}$ $6W$) & 0.01 & 0.2 & 
1.6 & 6.5 & 14.6 & 17.6 & 8.8 \\ \hline
BR($t^{\prime }\bar{b}^{\prime }/\bar{t}^{\prime }b^{\prime }\rightarrow b%
\bar{b}$ $5W)$ & - & 0.05 & 0.8 & 4.9 & 14.6 & 21.9 & 13.2 \\ \hline
BR($b^{\prime }\bar{b}^{\prime }\rightarrow b\bar{b}$ $4W)$ & - & - & 0.2 & 
2.9 & 13.2 & 26.3 & 19.8 \\ \hline
BR($t^{\prime }\bar{b}^{\prime }/\bar{t}^{\prime }b^{\prime }\rightarrow b%
\bar{b}$ $3W)$ & - & - & - & 1.1 & 9.9 & 29.6 & 29.6 \\ \hline
BR($t^{\prime }\bar{t}^{\prime }\rightarrow b\bar{b}$ $2W)$ & - & - & - & -
& 4.9 & 29.6 & 44.4 \\ \hline
\end{tabular}

\bigskip 

Table\ 7 : Branching Ratios (in \%) as a function of the number of
leptons ($l=e,\mu $) for the various final states, under the assumptions BR$(t^{\prime
}\rightarrow b^{\prime }W)\gg $ BR$(t^{\prime }\rightarrow bW)\ $and 
BR$(b^{\prime }\rightarrow t^{\prime }W)\gg $ BR$(b^{\prime }\rightarrow tW)$.

\newpage 

\section{Summary}

In this work, we reconsidered the available parameter space for a perturbative fourth generation of fermions, as allowed by the current bounds, namely direct and from Higgs searches, as well as by the electroweak precision observables. Our results show that the relation (\ref{1}) overconstrains the allowed SM4 scenarios if a small but non-zero CKM mixing between the two heaviest quark families is
required. Considering the bound (\ref{2}) and accounting for large mass splittings, we find that the heavy-to-heavy transitions $t^{\prime }\rightarrow Wb^{\prime }$\ ($b^{\prime
}\rightarrow t^{\prime }W$) provide relevant decay modes if the 3-4 family CKM
mixing element is smaller than 0.1. Additionally, we confirm that the $Z\rightarrow b\bar{b}$ decay rate gives a stronger bound on the 3-4 family CKM mixing element than the oblique parameters, independently of the Higgs and lepton sectors, and such that the $t^{\prime
}\rightarrow b W$ and $b^{\prime }\rightarrow t W$ decays are significantly suppressed.  

\bigskip

The tantalizing hints of a light Higgs boson at the LHC in the $\gamma \gamma $ channel, if confirmed, would exclude a fourth generation of fermions. However, the large suppression of the $H \gamma \gamma $ coupling due to the novel large Yukawa couplings casts some doubts on the perturbative validity of the NLO electroweak corrections. With more statistics, the comparison with the SM expected rates in the modes $H\rightarrow WW^{(\ast )}$, $H\rightarrow ZZ^{(\ast )}$ and $%
H\rightarrow f\bar{f}$ should allow to conclude whether the confirmed evidence of a light Higgs boson is incompatible with a fourth generation. While the Higgs mass range 120-600 GeV is claimed to be excluded below the two sigma level from the recent CMS and ATLAS results, we argue that $m_{H}\geq 450$ GeV requires more scrutiny given the large width of the Higgs boson in this range. Given the importance of excluding a larger spectrum from the direct searches, we have presented scenarios which could have evaded the experimental analyses. Even if $t^{\prime }\bar{t}^{\prime }$ ($b^{\prime }\bar{b}^{\prime }$)
pair production is unaffected by the magnitude of the fourth generation mixings, the possible importance of the heavy-to-heavy transitions $t^{\prime }\rightarrow b^{\prime }$ and $%
b^{\prime }\rightarrow t^{\prime }$ opens new challenging channels for the forthcoming
searches if a very small 3-4 family CKM mixing element is allowed. For extremely small values of this 3-4 mixing element, long-lived quarks and possibly bound states evade the direct searches, and require dedicated analyses \cite{46}.  

\bigskip

In particular, the
$t^{\prime }\bar{b}^{\prime }$\ and $\bar{t}^{\prime }b^{\prime
} $ electroweak productions in the $t$-channel can overcome $t^{\prime }b$ and $%
tb^{\prime }$ single production, and be accessed in the forthcoming searches. Allowing for signals with a higher jet multiplicity and same-sign dileptons, the subsequent decays can provide a clean signature against the SM background in all corresponding final states. These scenarios remain crucial to settle the question whether a perturbative and sequential fourth generation can be ruled out from direct searches at the LHC. In the case where the fourth generation couplings to the lighter families are small, we suggest that search strategies at the LHC should include both strong and weak production with multiple $W$ final state signatures.

\newpage 

\emph{Note added:} While this work was being completed, a related analysis by Dighe \emph{et al.} has appeared in \cite{50}, in which values of $m_{H}$ as large as $800$ GeV are considered. While similar results are obtained in this particular case, we also give a more specific focus on the light Higgs scenario, taking into account the current bounds for $m_{H}=125$ GeV. In this particular case, representative benchmark scenarios are provided from a systematic scan of the SM4 parameter space, consistently with the electroweak precision constraints. Assuming a small 3-4 CKM mixing, we discuss the corresponding production modes and decay signatures at the LHC, and highlight the requirement for new search strategies. 

\bigskip

\textbf{Acknowledgements}

The work of M.B. is supported by the National Fund for Scientific Research
(F.R.S.- FNRS, Belgium) under a FRIA grant. The authors thank Jorgen
D'Hondt, Petra Van Mulders and the members of the IIHE (Vrije Universiteit
Brussel) for the enjoyable collaboration on this project. M.B.\ would like to
acknowledge George W.S. Hou and Kai-Feng Chen for their warm hospitality, as well as the participants of the \textquotedblleft Focus Workshop on Heavy
Quarks at LHC\textquotedblright\ in Taipei (National Taiwan University).

\end{document}